\documentclass[mypaper,7pt,twoside]{CoAst}
\usepackage{epsf,graphicx,fancyhdr}
\renewcommand{\chaptermark}[1]{\markboth{#1}{}}

\newcommand{\aap}{A\&A}  %*************Stimmts?

\newcommand{\kopf}{\small\itshape Comm. in Asteroseismology \\ Contribution to the Proceedings of the 38$^{th}$\,LIAC\,/\,HELAS-ESTA\,/\,BAG, 2008
}

\newcommand{\Authors}[1]{\begin{center}\normalsize\bf\sf #1 \end{center}}

\renewcommand{\author}[1]{\begin{center}\normalsize\bf\sf #1 \end{center}}
\newcommand{\Address}[1]{\begin{center}\small\sf #1 \end{center}}

\DeclareGraphicsExtensions{.eps,.jpg}

\newcommand{\Session}[1]{{\vspace{3mm}\small \noindent  \hspace*{3mm} Session: } #1 \normalsize}

\newcommand{\Objects}[1]{{\vspace{0mm}\small \noindent  \hspace*{3mm} Individual Objects: } \small #1 \normalsize}

	\newcommand{\four}{\small Observed frequencies in pulsating massive stars}

\renewenvironment{abstract}{\section*{Abstract}\normalsize\sf}{}
\newcommand{\References}[1]{\begin{flushleft}{\large References\\}\vspace*{2mm}\small #1 \end{flushleft}}

\newcommand{\chapterCoAst}[2]{\chapter[\sf\normalsize #1\\ \footnotesize \hspace*{5mm}by #2 \sf\normalsize][]{#1\\}\rhead[\fancyplain{}{\sf\footnotesize \center{#1}}]{\fancyplain{}{\sffamily\thepage}}\lhead[\fancyplain{\kopf}{\sffamily\thepage}]{\fancyplain{\kopf}{\sf\footnotesize \center{#2}}}}

\renewcommand{\chaptername}{}

\newcommand{\acknowledgments}[1]{\vspace*{5mm}\noindent  \textbf{Acknowledgments.} #1}

\def\rfr{\smallskip\par\noindent
        \hangindent=7truemm
        \hangafter=1}
\begin{document}
\sf
\chapterCoAst{Massive $B$-type pulsators in low-metallicity environments}{Karoff et al.}
\Authors{C.\,Karoff$^{1,2}$, T.\,Arentoft$^2$, L.\,Glowienka$^2$, C.\,Coutures$^3$, T.\,B.\,Nielsen$^2$,  G.\,Dogan$^{2}$, F.\,Grundahl$^2$, and H.\,Kjeldsen$^2$} 
\Address{
$^1$ School of Physics and Astronomy, University of Birmingham, Edgbaston, Birmingham B15 2TT, UK\\
$^2$ Danish AsteroSeismology Centre (DASC), Department of Physics and Astronomy, University of Aarhus, DK-8000 Aarhus C, Denmark \\
$^3$ Institut d'Astrophysique de Paris, CNRS, Universite Pierre et Marie Curie UMR7095,
 98bis  Boulevard Arago, 75014 Paris, France\\
}
\noindent
\begin{abstract}
Massive $B$-type pulsators such as $\beta$ Cep and slowly pulsating B (SPB) stars pulsate due to layers of increased opacity caused by partial ionization. The increased opacity blocks the energy flux to the surface of the stars which causes the layers to rise and the opacity to drop. This cyclical behavior makes the star act as a heat engine and the star will thus pulsate. For $\beta$ Cep and SPB stars the increased opacity is believed to be caused by partial ionization of iron and these stars should therefore contain non-insignificant quantities of the metal. A good test of this theory is to search for $\beta$ Cep and SPB stars in low-metallicity environments. If no stars are found the theory is supported, but on the other hand if a substantial number of $\beta$ Cep and SPB stars are found in these environments then the theory is not supported and a solutions needed. With a growing number of identified  $\beta$ Cep and SPB stars in the low-metallicity Magellanic Clouds we seem to be left with the second case. We will in this context discuss recent findings of $\beta$ Cep and SPB stars in the Magellanic Clouds and some possible solutions to the discrepancy between these observations and the theory. We also describe an ambitious project that we have initiated on the Small Magellanic Cloud open cluster NGC 371 which will help to evaluate these solutions.
\end{abstract}

\Session{ \four } \\% you can chose from, \one, \two, ... \six, \ESTA, \future or \poster
\Objects{NGC 371} 
\section*{The Problem}
Though the number of $\beta$ Cep and SPB stars is predicted to be very limited in the Magellanic Clouds due to the reduced metallicity in these environments a growing number of studies are identifying $\beta$ Cep and SPB in the Magellanic Clouds.  Pigulski \& Ko{\l}aczkowski (2002) used OGLE II and MACHO data to identify 3 $\beta$ Cep in the Large Magellanic Cloud (LMC). This study was later updated by Ko{\l}aczkowski et al. (2006) who identified 92 $\beta$ Cep and 6 SPB stars in the LMC and  59 $\beta$ Cep and 11 SPB stars in the Small Magellanic Cloud (SMC). Diago et al. (2008) have reanalyzed MACHO data of 186 absorption-line B stars and identified 1 $\beta$ Cep and 8 SPB stars in the SMC. Recently, Karoff et al. (2008) have identified 29 candidate SPB stars in the young open SMC cluster NGC 371. This result is particularly interesting because (if confirmed) it indicates that the population fraction of SPB stars in this cluster is probably larger than or equal to the population fraction of SBP stars in the Galaxy. 

The metallicity of Magellanic Clouds is found to be around $Z$ = 0.007 and $Z$ = 0.002 for the LMC and SMC, respectively (Maeder et al., 1999) . Though no instability domains have been published for such low metallicities there are indications that pulsation can only be driven down to 0.01 in $\beta$ Cep stars and down to 0.005 in SPB stars using standard physics (solar abundances from Grevesse \& Noels [1993]) which means that the increasing number of observed $\beta$ Cep and SPB stars in the Magellanic Clouds cannot be explained by standard stellar models. Pamyatnykh (1999) has calculated instability domains down to $Z$ = 0.01 and Miglio et al. [2007a,b] have calculated instability domains down to $Z$ = 0.005 where the SPB domain seems to have vanished, but as no domains were calculated for lower metallicities it cannot be concluded that the SPB domain disappears for lower metallicities. There is clearly an urgent need for calculations of instability domains of SPB stars at the metallicity of the SMC ($Z$ = 0.002), as it has not been proven that pulsation in SPB stars cannot be driven at this metallicity using standard stellar models, though it seems doubtful.
\newpage

\section*{The Solutions} 
Though the discrepancy between the lowest metallicity at which standard stellar models predict pulsation in $\beta$ Cep and SPB stars and the metallicity of the $\beta$~Cep and SPB stars in especially the SMC is quite large we can identify three possible extensions to the standard models which can make the low-metallicity $\beta$ Cep and SPB stars pulsate:

\begin{itemize}
\item The stars that pulsate have higher metallicity than the average metallicity of the Magellanic Clouds.
\item The new (solar) abundances might explain why pulsation can be driven at lower metallicity.
\item Local iron enhancement might drive pulsation at lower metallicity in these stars.
\end{itemize}

\subsection*{Metallicity}
Of course not all the stars in the Magellanic Clouds have the same metallicity and as indicated by the wide range of published estimates of the metallicity ($Z$ = 0.004 to 0.01 for LMC and 0.001 to 0.003 for SMC [Maeder et al., 1999]) there are differences in the values of the metallicities within the two clouds. This probably reflects the fact that the Magellanic Clouds are still subject to a reasonable amounts of star formation, which is also seen in the large correlation between age and metallicity for the two clouds (see e.g. Pagel \& Tautvai{\v s}ien{\.e}, [1999]). Another important issue to remember when using the total metallicity $Z$ (defined as the mass fraction of heavy elements to hydrogen) to evaluate a star's ability to drive pulsation is that $Z$ is not a direct measurement of the iron content in the star. In fact the most important elements for calculating $Z$ are O, C and Ne and then Fe. This means that a low $Z$ does not necessarily reflect a low iron content. It could also reflect e.g. a low oxygen content. The problem is of course that we only have detailed abundance estimates for a limited number of stars in the Magellanic Clouds and that standard stellar models are calculated using the total metallicity $Z$, and then assuming solar abundances instead of using the individual abundances as e.g. spectroscopic metallicity [Fe/H].

\subsection*{The New Solar Abundances}
In this way the abundance of the Sun becomes important for the ability of $\beta$ Cep and SPB stars in the Magellanic Clouds to drive pulsation. Miglio et al. (2007a,b) have shown that the instability strip in the Hertzsprung-Russell diagram is increased for both $\beta$ Cep and SPB stars, by using the new abundances of Asplund et al. (2005). Especially it is shown by Miglio et al. (2007a,b) that the new abundances and the new opacities can drive radial oscillations in stars with a metallicity as low as $Z$ = 0.01, which cannot be done with the old abundances and opacities. It is also seen that the new abundances have the effect of extending the excited frequencies towards higher overtones. 

Though there are contradictory interpretations of the effect of the new abundances on the excitations of $\beta$ Cep and SPB stars (see for example Pamyatnykh \& Ziomek, 2007) it might be possible to test the new abundances and opacities on oscillations in low-metallicity $\beta$ Cep and SPB stars. 

\subsection*{Local Iron Enhancement}
The last solution to the problem with the low-metallicity massive $B$-type pulsators is the same solution that was found to solve the problem of pulsating subdwarf B (sdB) stars. To begin with, the oscillations in these stars were believed to be driven by the He$_{\mathrm{II}}$ -- He$_{\mathrm{III}}$ convection zone; however, it was soon realized that the driving was negligible in the He$_{\mathrm{II}}$ -- He$_{\mathrm{III}}$ convection zone as this region contains only very little mass, and therefore it carries practically no inertia to drive the pulsations. Instead it was shown by Charpinet et al. (1996, 1997) that pulsation in sdB stars could be driven by an opacity bump due to a local iron enhancement in the envelope of these stars. The enhancement was shown to be caused by gravitational settling and radiative levitation of heavy elements. A similar mechanism for $\beta$ Cep and SPB stars was suggested by Pamyatnykh et al. (2004) and has been investigated for $\beta$ Cep and SPB stars in low-metallicity environments by Bourge et al., (2006, 2007),  Bourge \& Alecian (2006) and Miglio et al. (2007c). Though it is still not clear if this is the solution to the possible low-metallicity $\beta$ Cep and SPB stars, one way to test this would be to look for chemical peculiarities at the surfaces of these stars, especially Si enrichment, which could reflect local enhancement of different elements (i.e. iron) by diffusion processes if the stars are not fast rotators.

\section*{Case Study -- NGC 371}
In order to test the prediction that $\beta$ Cep and SPB stars should be rare in the Magellanic Clouds we have initiated an ambitious project which includes: 1) a survey for candidate $\beta$ Cep and SPB stars in the open SMC cluster NGC 371; 2) spectroscopic determination of important physical parameters; 3) determination of precise eigenmode frequencies based on a multi-site campaign and; 4) detailed modeling of the stars. The result of the first part of this project was the discovery of 29 candidate SPB stars in the cluster (amplitude spectra for the 29 candidates are shown in Fig.~1; see Karoff et al. [2008] for details), a discovery that clearly contradicts the theoretical predictions. The next step is therefore to obtain high-resolution spectra for these stars in order 1) to discriminate between binaries and bona fide SPB stars; 2) to determine cluster membership and; 3) to obtain important physical parameters of the candidates. This step in now underway as stellar spectra are now being obtained from the Gemini Multi-Object Spectrographs ($R = 4~400$) as part of variable star one-shot project (Dall et al. 2007) We have also applied for VLT time with the Fibre Large Array Multi Element Spectrograph ($R = 18~470$) for this cluster.

\acknowledgments{
CK acknowledges financial support from the Danish Natural Science Research Council and CK, FG, GD and TA also acknowledge support from the Danish AsteroSeismology Centre. 
}
\References{
\rfr Asplund, M., Grevesse, N., Sauval, A. J., et al. 2005, A\&A 431, 693 
\rfr Charpinet, S., Fontaine, G., Brassard, P., et al., 1996, ApJ, 471, L103
\rfr Charpinet, S., Fontaine, G. \& Brassard, P., et al. 1997, ApJ, 483, L123
\rfr Cox A. N., Morgan S. M., Rogers F. J. \& Iglesias C. A., 1992, ApJ, 393, 272
\rfr Basu, S. \& Antia, H. M., 2008, Phys. Rep., 457, 217
\rfr Bourge, P.-O., Th\'eado, S. \& Thoul, A., 2007, Commun. Asteroseismol., 150, 203
\rfr Bourge, P.-O., Alecian, G., Thoul, A., et al., 2006, Commun. Asteroseismol., 147, 105
\rfr Bourge, P.-O. \& Alecian, G., 2006, ASP, 249, 201
\rfr Dall, T. H., Foellmi, C., Pritchard, J., et al., 2007, \aap, 470, 1201
\rfr Diago, P. D., GutiŽrrez-Soto, J., Fabregat, J., et al., 2008, A\&A, 480, 179
\rfr Grevesse, N., \& Noels, A., 1993, Origin and Evolution of the Elements, 15 
\rfr Karoff C., Arentoft T., Glowienka L., 2008, MNRAS, 386, 1085
\rfr Ko{\l}aczkowski, Z., Pigulski, A., Soszy{\'n}ski, I., et al., 2006, Mem. Soc. Astron. Ital., 77, 336
\rfr Maeder, A., Grebel, E.~K., \& Mermilliod, J.-C.\ 1999, A\&A, 346, 45
\rfr Miglio A., Montalb \& J., Dupret M.-A., 2007a, MNRAS, 375, L21 
\rfr Miglio A., Montalb \& J., Dupret M.-A., 2007b, Commun. Asteroseismol., 151, 48 
\rfr Miglio A., Bourge P.-O., Montalb \& J., Dupret M.-A., 2007c, Commun. Asteroseismol., 150, 209 
\rfr Pamyatnykh, A. A. 1999, MNRAS, 49, 119
\rfr Pamyatnykh, A. A., Handler, G. \& Dziembowski, W. A., 2004, MNRAS, 350, 1022
\rfr Pamyatnykh, A. A. \& Ziomek, W. 2007, Commun. Asteroseismol., 150, 207 
\rfr Pigulski A. \& Ko{\l}aczkowski, Z., 2002, A\&A, 388, 88
\rfr Pagel, B.~E.~J. \& Tautvai{\v s}ien{\.e}, G.\ 1999, Ap\&SS, 265, 461
}

\begin{figure}[!h]
	\centering
          \includegraphics[width=3.7cm]{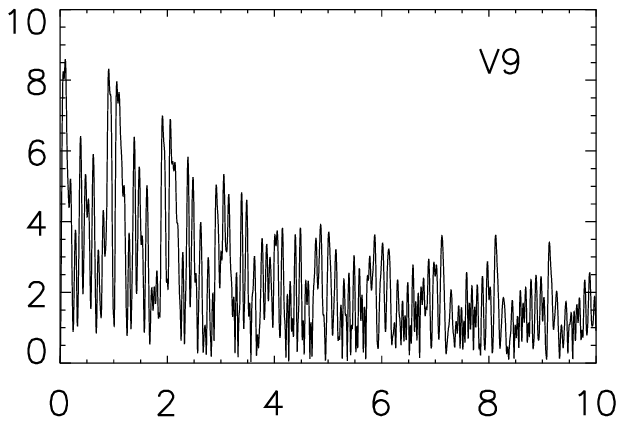}
	 \includegraphics[width=3.7cm]{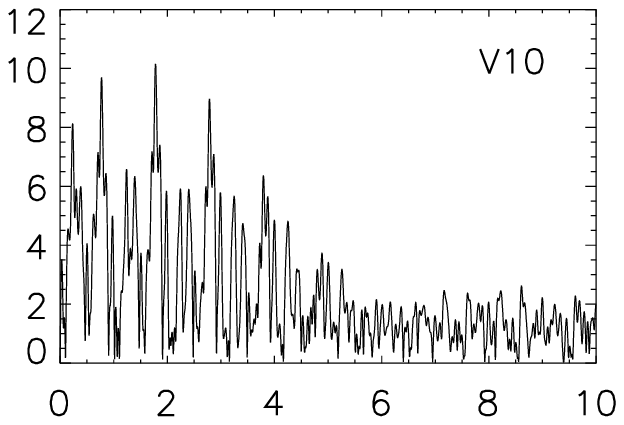}
	 \includegraphics[width=3.7cm]{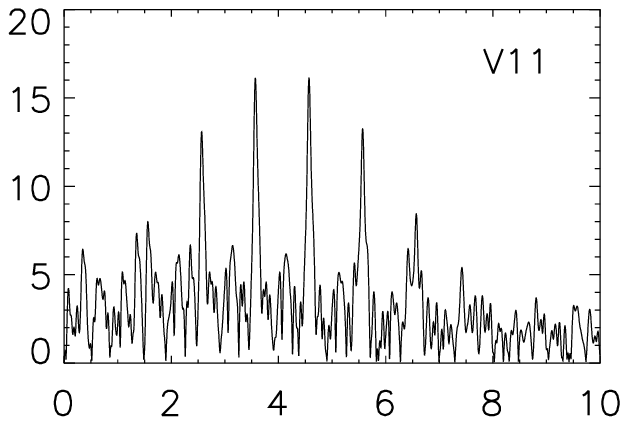}
	 \includegraphics[width=3.7cm]{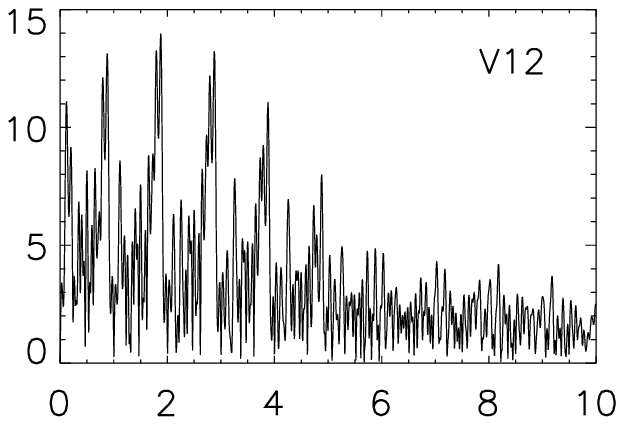}
	 \includegraphics[width=3.7cm]{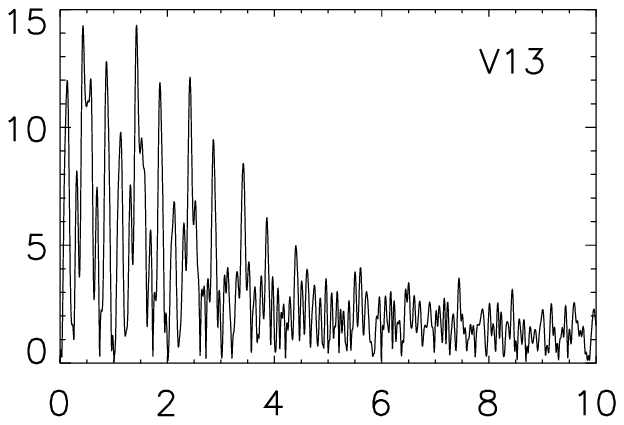}
	 \includegraphics[width=3.7cm]{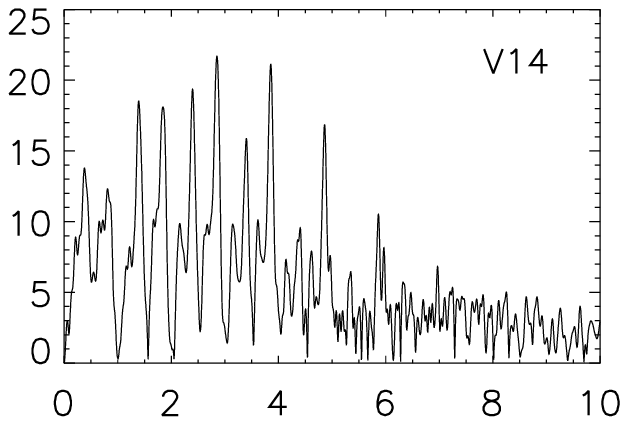}
	 \includegraphics[width=3.7cm]{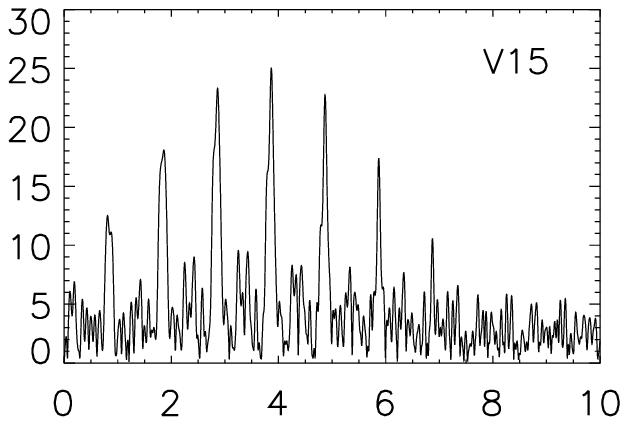}
	 \includegraphics[width=3.7cm]{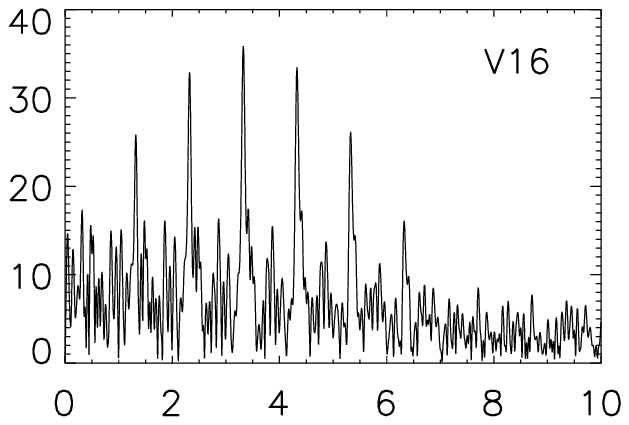}
	 \includegraphics[width=3.7cm]{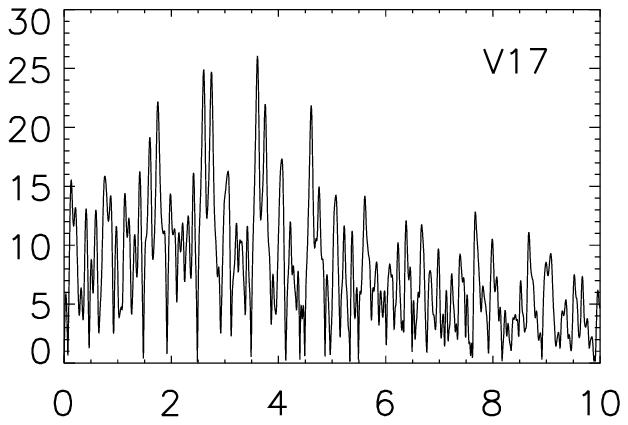}
	 \includegraphics[width=3.7cm]{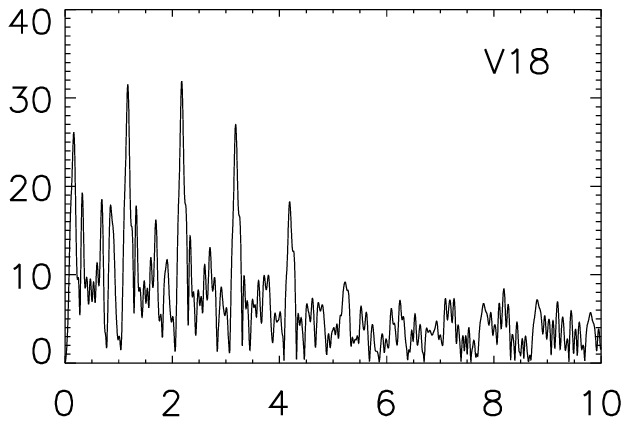}
	 \includegraphics[width=3.7cm]{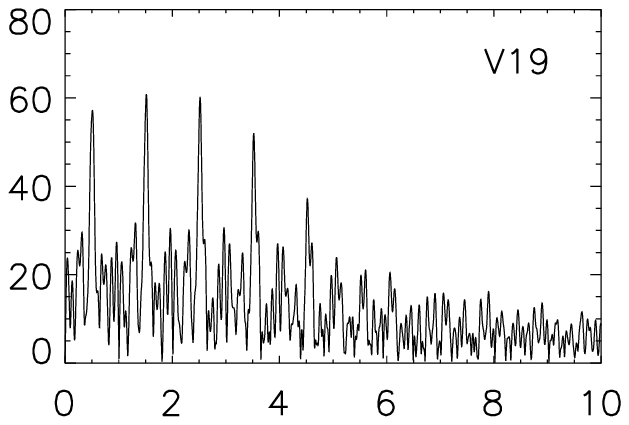}
	 \includegraphics[width=3.7cm]{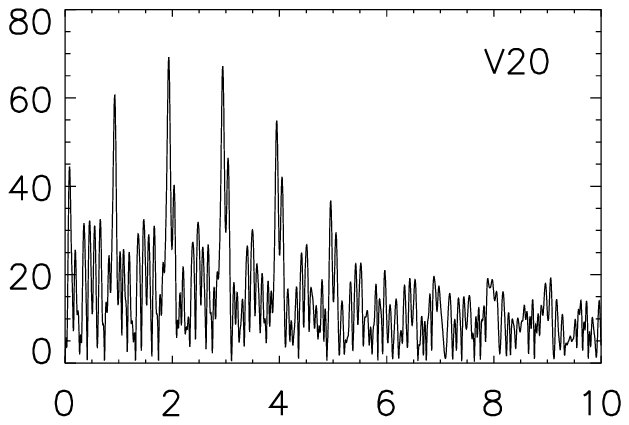}
	 \includegraphics[width=3.7cm]{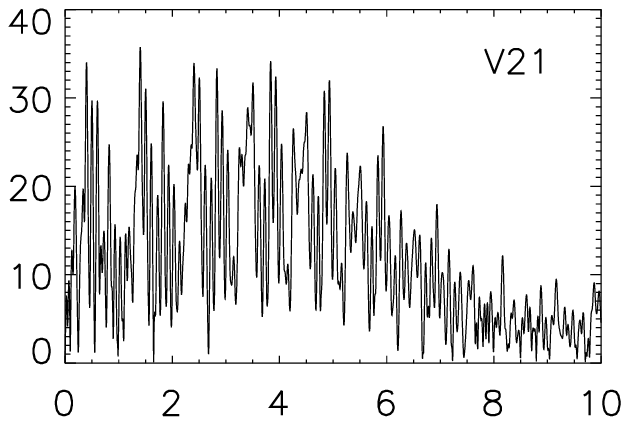}
	 \includegraphics[width=3.7cm]{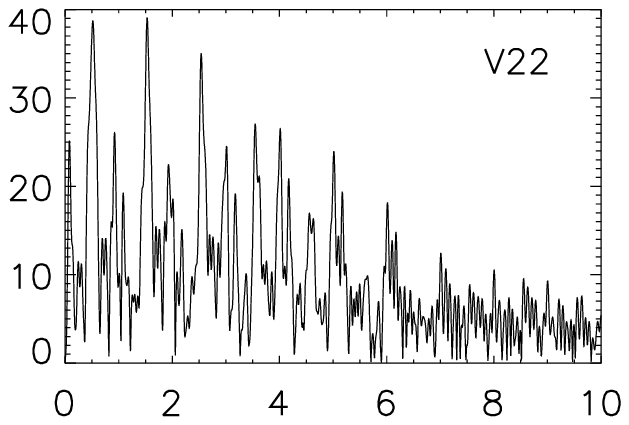}
	 \includegraphics[width=3.7cm]{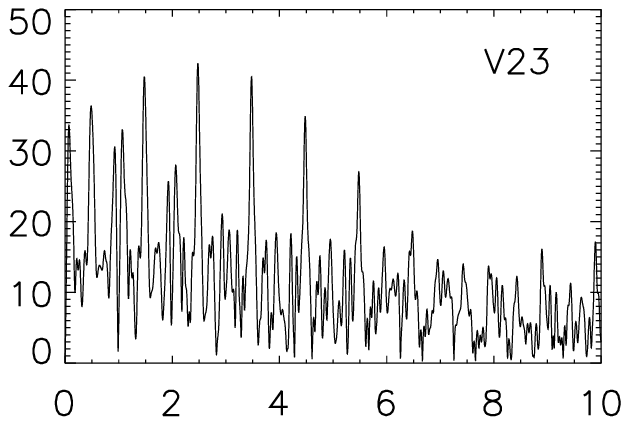}

	\caption{Amplitude spectra of the 27 stars pulsating in the upper part of the main sequence of NGC 371. The y-axes are in mmag (in $B$) and the x-axes are in c/d. Note the different scaling of the y-axes.}
          \label{lable}
\end{figure}
\addtocounter{figure}{-1}
\begin{figure}[!h]
	\centering
		 \includegraphics[width=3.7cm]{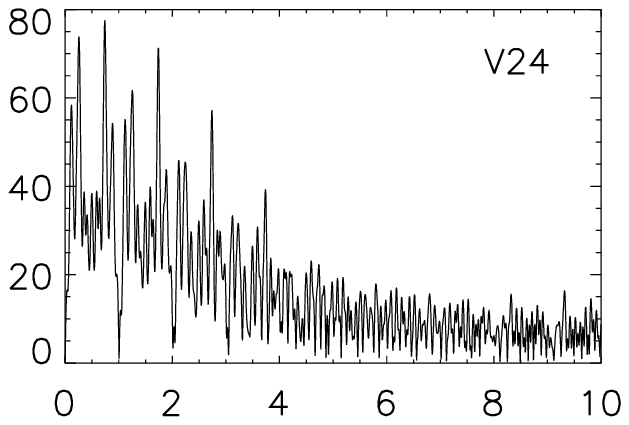}
	 \includegraphics[width=3.7cm]{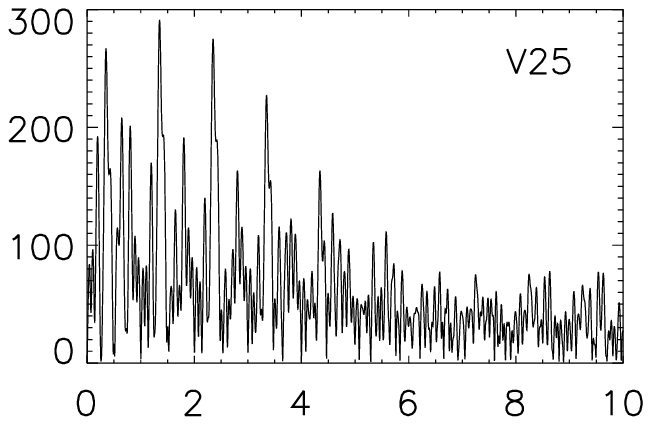}
	 \includegraphics[width=3.7cm]{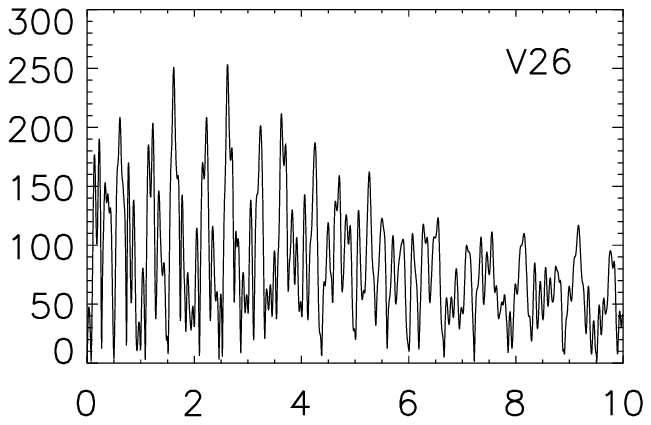}
		 \includegraphics[width=3.7cm]{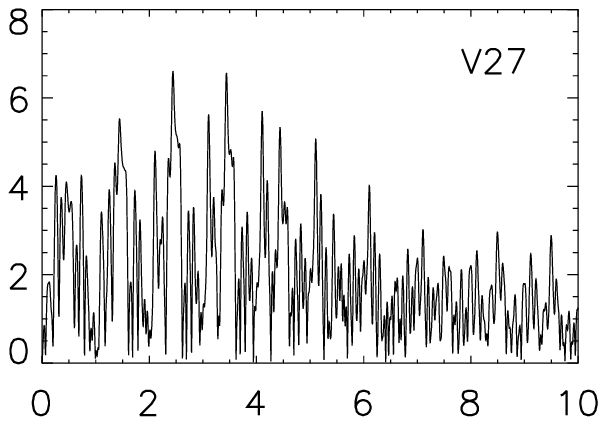}
	 \includegraphics[width=3.7cm]{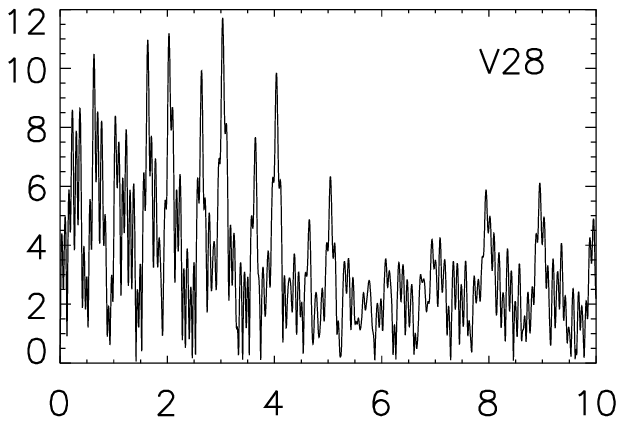}
	 \includegraphics[width=3.7cm]{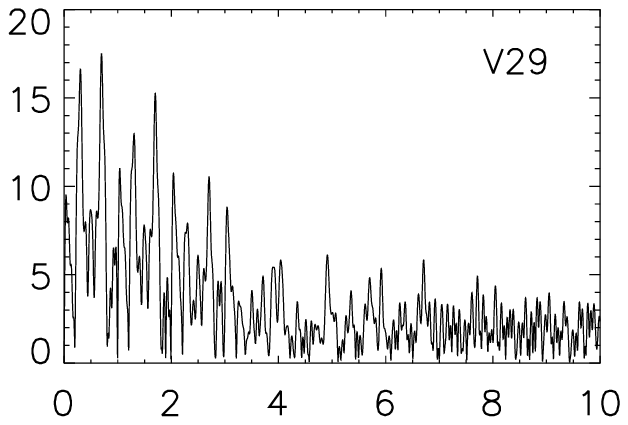}
	 \includegraphics[width=3.7cm]{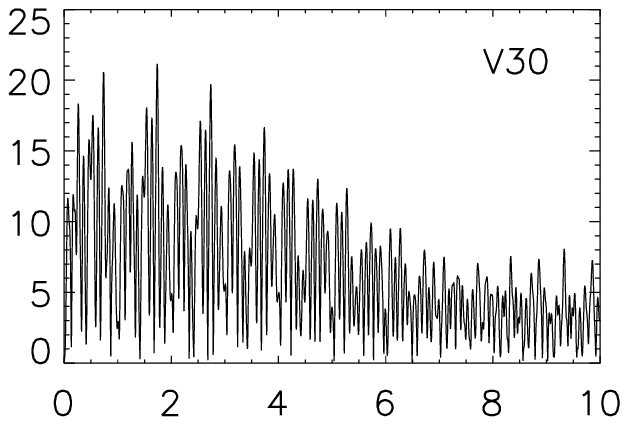}
	 \includegraphics[width=3.7cm]{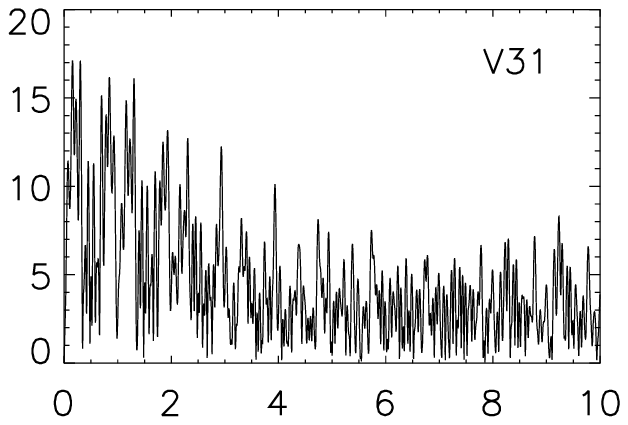}
	 \includegraphics[width=3.7cm]{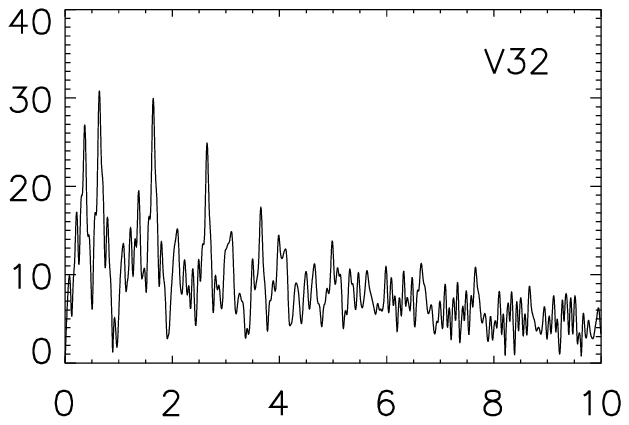}
	 \includegraphics[width=3.7cm]{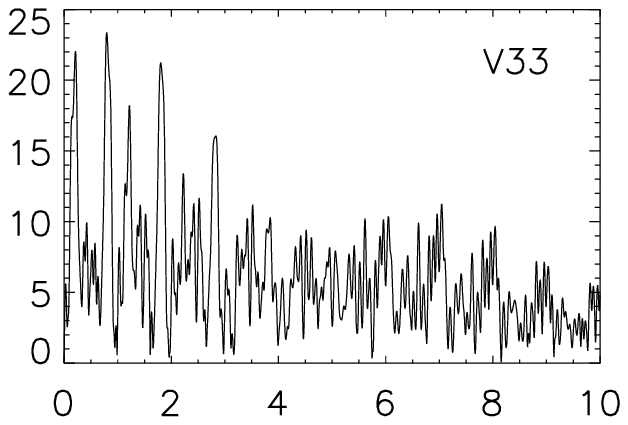}
	 \includegraphics[width=3.7cm]{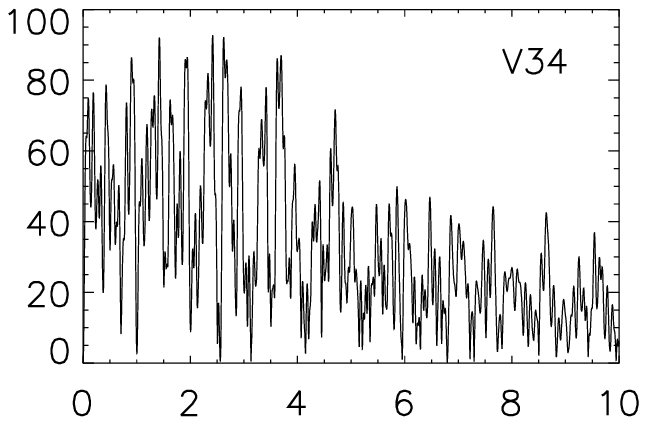}
	 \includegraphics[width=3.7cm]{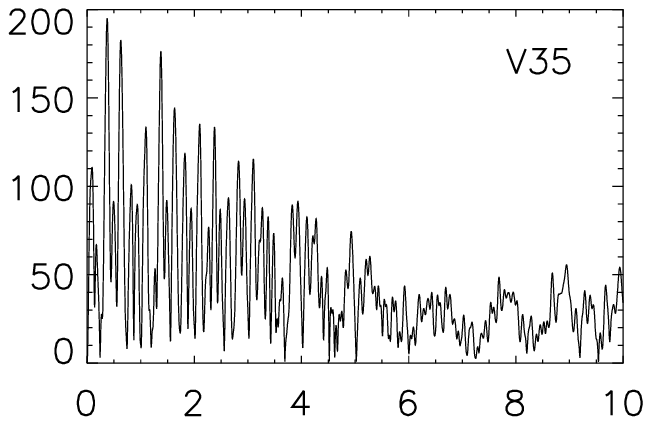}
	 \includegraphics[width=3.7cm]{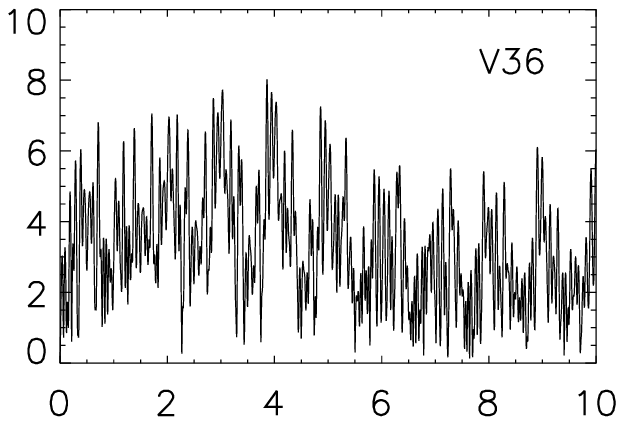}
	 \includegraphics[width=3.7cm]{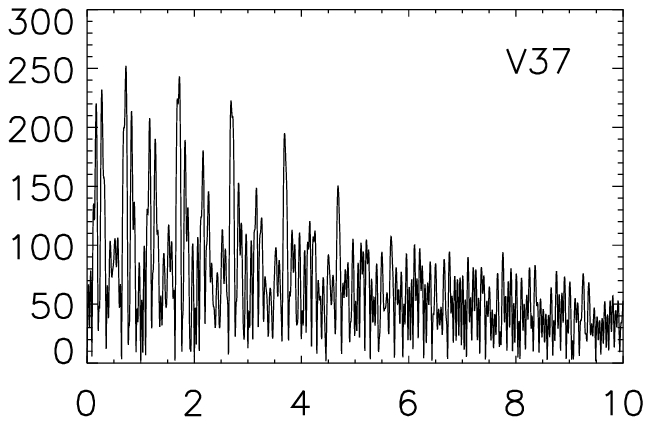}
	\caption{-- continued}
          \label{lable}
\end{figure}

\end{document}